\documentclass{ws-procs9x6}

\newcommand{\nn}{\nonumber}
\newcommand{\be}{\begin{equation}}
\newcommand{\ee}{\end{equation}}
\newcommand{\ba}{\begin{eqnarray}}
\newcommand{\ea}{\end{eqnarray}}

\def\gev{~{\rm GeV}}
\def\ale{\alpha_{\rm elm}}

\newcommand{\lsim}{\raisebox{-4pt}{$\,\stackrel{\textstyle
                                                         <}{\sim}\,$}}

\newcommand{\ov}[1]{\overline#1}

\begin{document}
\title{ GPDs, form factors and Compton scattering}

\author{P.\ Kroll}

\address{Fachbereich Physik, Universit\"at Wuppertal,\\ 
D-42097 Wuppertal, Germany\\
Email: kroll@physik.uni-wuppertal.de}

\maketitle

\abstracts{
The basic theoretical ideas of the handbag factorization and its
application to wide-angle scattering reactions are reviewed. With
regard to the present experimental program carried out at JLab, 
wide-angle Compton scattering is discussed in some detail.} 
\section{Introduction}
As is well-known factorization is an important property of QCD
whithout which we would not be able to calculate form factors or cross
sections. Factorization into a hard parton-level subprocess to be
calculated from perturbative QED and/or QCD, and soft hadronic matrix
elements which are subject to non-perturbative QCD and are not
calculable at present, has been shown to hold for a number of reactions 
provided a large scale, i.e.\ a large momentum transfer, is available. 
For other reactions factorization is a reasonable hypothesis. In the 
absence of a large scale we don't know how to apply QCD and, for the 
interpretation of scattering reactions, we have to rely upon effective 
theories or phenomenological models as for instance the Regge pole one.  

\begin{figure}[t]
\begin{center}
\includegraphics[width=3.8cm,bbllx=45pt,bblly=230pt,bburx=545pt,
bbury=615pt,clip=true]{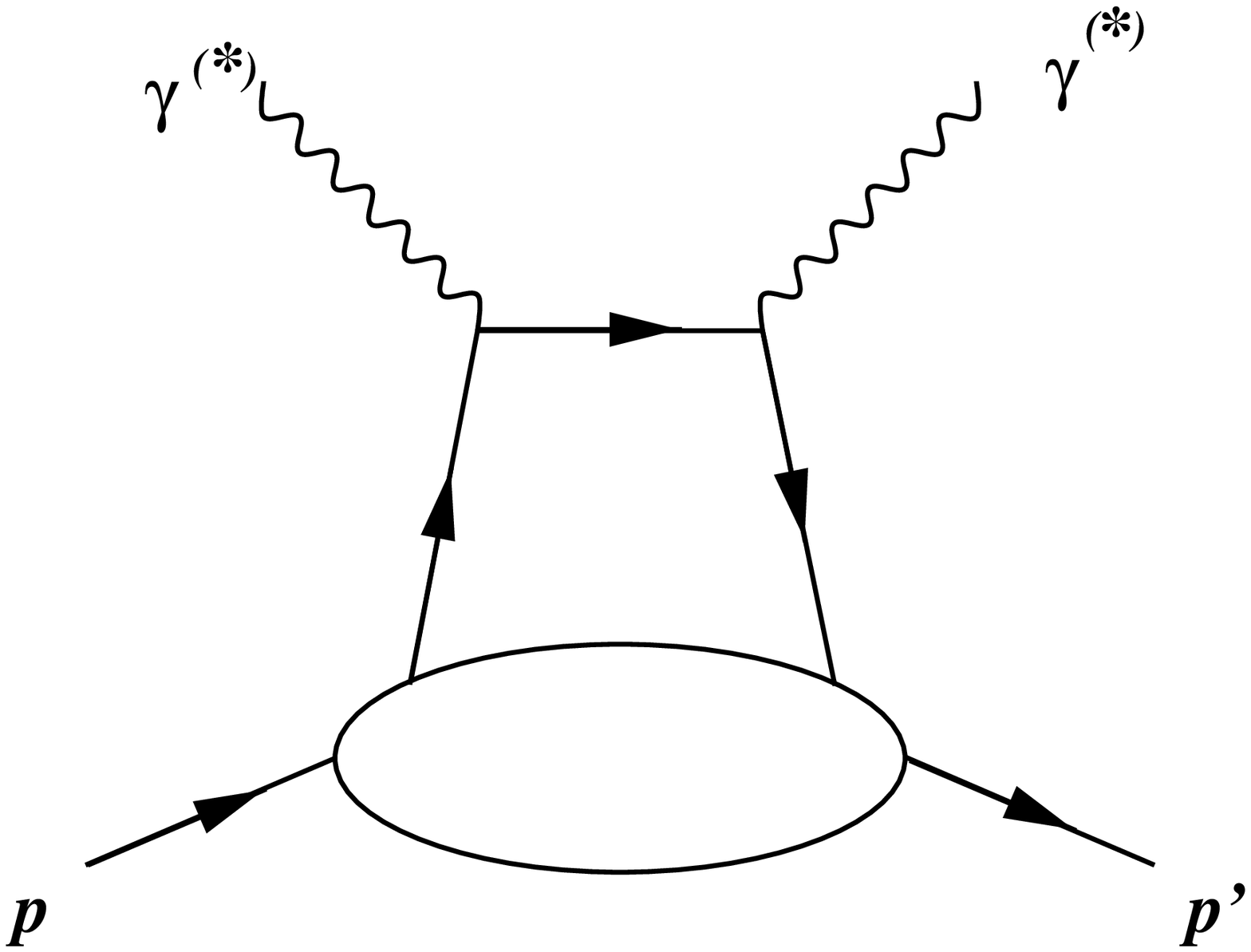} 
\includegraphics[width=4.5cm,bbllx=73pt,bblly=525pt,bburx=510pt,
bbury=795pt,clip=true]{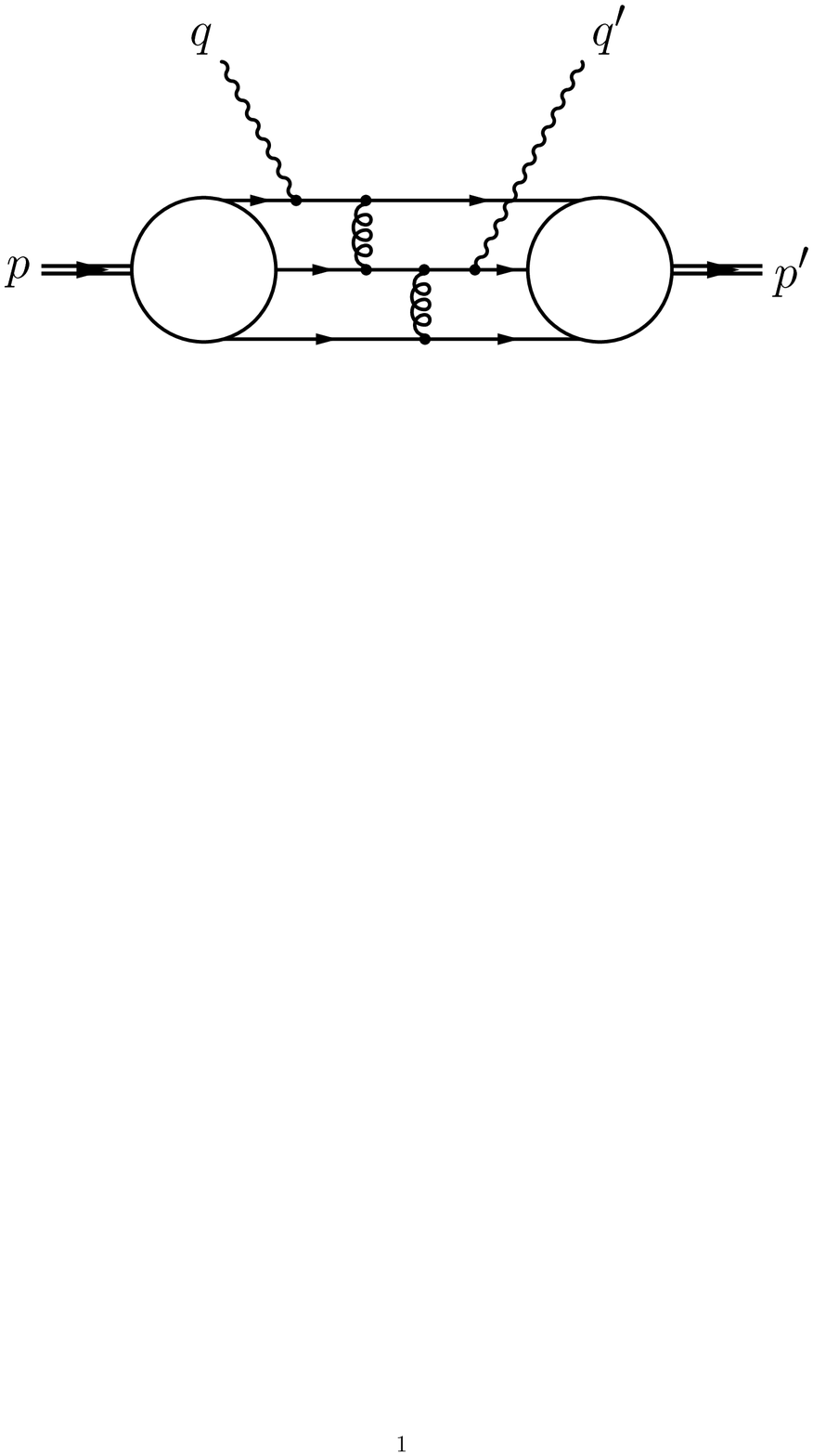} 
\caption{{}Handbag (left) and leading-twist (right) factorization for
Compton scattering.}
\label{fig:handbag}
\end{center}
\end{figure}

For hard exclusive processes there are two different factorization
schemes available. One of the schemes is the handbag factorization
(see Fig.\ \ref{fig:handbag}) where only one parton participates in 
the hard subprocess (e.g.\ $\gamma q\to \gamma q$ in Compton scattering) 
and the soft physics is encoded in generalized parton distributions
(GPDs) \cite{mue1994,rad97}. The handbag approach applies to deep
virtual exclusive scattering (e.g.\ DVCS) where one of the photons has
a large virtuality, $Q^2$, while the squared invariant momentum
transfer, $-t$, from the ingoing hadron to the outgoing one is
small. It also applies to wide-angle scattering (WACS) where $Q^2$ is
small while $-t$ (and $-u$) are large \cite{rad98,DFJK1}. This class
of reactions is the subject of my talk. For wide-angle scattering
there is an alternative scheme, the leading-twist factorization
\cite{bro80}. Here all valence quarks the involved hadrons are made
off participate in the hard scattering (e.g.\ $\gamma qqq\to \gamma
qqq$ in Compton scattering) while the soft physics is encoded in
distribution amplitudes representing the probability amplitudes for
finding quarks in a hadron with a given momentum distribution (see
Fig.\ \ref{fig:handbag}).    
Since neither the GPDs nor the distribution amplitudes can be
calculated whithin QCD at present, it is difficult to decide
which of the factorization schemes provides an appropriate description 
of, say, wide-angle Compton scattering at $-t\simeq 10 \gev^2$. The 
leading-twist factorization probably requires larger $-t$ than the
handbag one since more details of the hadrons have to be resolved. 
Recent phenomenological and theoretical developments \cite{BK} support
this conjecture. In the following I will discuss the handbag contribution 
only, assuming that the leading-twist one is negligibly small for
momentum transfers of about $10\gev^2$. The ultimate decision whether or 
not this assumption is correct, is to be made by experiment.

It should be noted that immediately after the discovery of the partons in
the late sixties constituent scattering models had been invented
\cite{bjo,sivers} which bear resemblance to the handbag contribution. As
compared to these early attempts the handbag factorization has now a
sound theoretical foundation. Particularly the invention of GPDs effectuated a 
decisive step towards a theoretical understanding of hard exclusive reactions.
\section{The handbag in wide-angle Compton scattering}
In 1998 Radyushkin \cite{rad98} calculated the handbag contribution to
Compton scattering starting from double distributions. Somewhat later Diehl,
Feldmann, Jakob and myself calculated it on the basis of parton
ideas \cite{DFJK1}. Both approaches arrived at essentially the same
results. Here, I will briefly describe our approach because I am more
familiar with it. Our kinematical requirements are that the three
Mandelstam variables $s,\, -t,\, -u$ are much larger than $\Lambda^2$
where $\Lambda$ is a typical hadronic scale of order $1\, \gev$. 
The bubble in the handbag is viewed as a
sum over all possible parton configurations as in deep ineleastic
lepton-proton scattering (DIS). The contribution we calculate is
defined by the requirement of restricted  parton virtualities,
$k_i^2<\Lambda^2$, and intrinsic transverse parton momenta, ${\bf
k_{\perp i}}$, which satisfy $k_{\perp i}^2/x_i <\Lambda^2$, where
$x_i$ is the momentum fraction parton $i$ carries. 

\begin{figure}[t]
\begin{center}
\includegraphics[width=2.3cm,bbllx=263pt,bblly=600pt,bburx=370pt, 
bbury=700pt,clip=true]{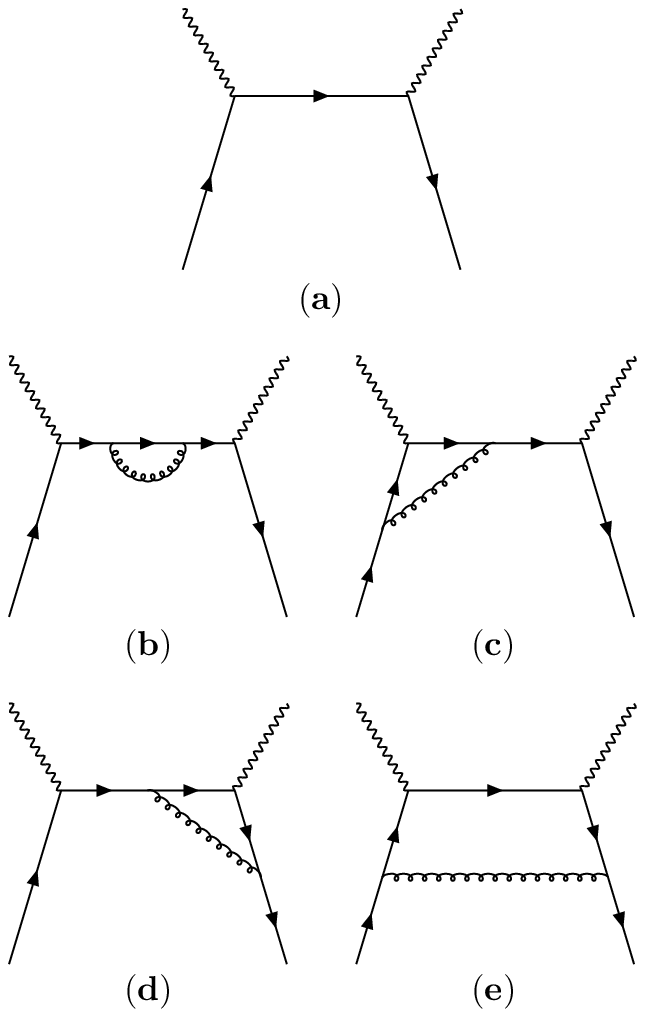}   
\includegraphics[width=3.5cm,bbllx=182pt,bblly=410pt,bburx=400pt, 
bbury=610pt,clip=true]{hkm-fig2.ps} 
\caption{Sample Feynman graphs for $\gamma q\to \gamma q$ to NLO in
perturbative QCD.}
\label{fig:NLO}
\end{center}
\end{figure}

It is of advantage to work in a symmetrical frame which is a c.m.s
rotated in such a way that the momenta of the incoming ($p$) and
outgoing ($p'$) proton momenta have the same light-cone plus components.  
In this frame the skewness, defined as
\be 
\xi = \frac{(p - p')^+}{(p + p')^+}\,,
\ee
is zero. One can then show that the subprocess Mandelstam variables
$\hat{s}$ and $\hat{u}$ are the same as the ones for the full process,
Compton scattering off protons, up to corrections of order
$\Lambda^2/t$:
\ba
\hat{s}=(k_j+q)^2 \simeq (p+q)^2 =s\,, \quad 
\hat{u}=(k_j-q')^2 \simeq (p-q')^2 =u\,.
\ea
The active partons, i.e.\ the ones to which the photons couple, are
approximately on-shell, move collinear with their parent hadrons and
carry a momentum fraction close to unity, $x_j, x_j' \simeq 1$.
Thus, like in DVCS, the physical situation is that of a hard
parton-level subprocess, $\gamma q\to \gamma q$, and a soft emission
and reabsorption of quarks from the proton. The helicity amplitudes
for WACS then read
\ba
\label{ampl}
{M}_{\mu'+,\,\mu +}(s,t) &=& \;2\pi\ale 
     \left[ { T}_{\mu'+,\,\mu+}(s,t)\,(R_V(t) + R_A(t))\,\right.\nn\\[0.5em]
&&\quad\left.  + \,  { T}_{\mu'-,\,\mu-}(s,t)\,(R_V(t) - R_A(t)) \right]  
                                                                \,,\\[0.5em]
 { M}_{\mu'-,\,\mu +}(s,t) &=& \;-\pi\ale \frac{\sqrt{-t}}{m} 
         \left[  T_{\mu'+,\,\mu+}(s,t)\, 
         + \,  { T}_{\mu'-,\,\mu-}(s,t)\, \right] \,R_T(t)\,.\nn
\ea
$\mu,\, \mu'$ denote the helicities of the incoming and outgoing
photons, respectively. The helicities of the protons in $ { M}$ and
quarks in  the hard scattering amplitude $ T$ are labeled by their
signs. The hard scattering has been calculated to next-to-leading order 
perturbative QCD \cite{hkm}, see Fig.\ \ref{fig:NLO}. To this order the 
gluonic subprocess, $\gamma g\to \gamma g$ has to be taken into account 
as well. The form factors $R_i$ represent $1/x$-moments of GPDs at
zero skewness.  
$R_T$ controls the proton helicity flip amplitude while the
combination $R_V+R_A$ is the response of the proton to the emission
and reabsorption of quarks with the same helicity as it and $R_V-R_A$
that one for opposite helicities. The identification of the form
factors with $1/x$-moments of GPDs is possible because the plus
components of the proton matrix elements dominate as in DIS and
DVCS. This is non-trivial feature given that, in contrast to DIS and
DVCS, not only the plus components of the proton momenta but also
their minus and transverse components are large here. A more technical
aspect is the fact that the handbag approach naturally demands the
use of light-cone techniques. Thus, (3) is a light-cone
helicity amplitude. To facilitate comparison with experiment one may 
transform the amplitudes (3) to the ordinary c.m.s.\ helicity
basis \cite{hkm,die01}. 
\section{Modeling the GPDs}
\label{sect:model}
The structure of the handbag amplitude, namely its representation as a
product of perturbatively calculable hard scattering amplitudes and
$t$-dependent form factors 
\be
  { M}(s,t) \sim  { T}(s,t)\, R(t)
\label{fact}
\ee
is the essential result. Refuting the handbag approach necessitates 
experimental evidence against the structure (\ref{fact}). In oder to 
make actual predictions for Compton scattering however models for the 
soft form factors or rather for the underlying GPDs are required.
A first attempt to parameterize the GPDs $H$ and $\widetilde{H}$  
at zero skewness reads \cite{rad98,DFJK1,hkm,hanwen} (see also
\cite{afanasev,bar93}) 
\ba
H^a(\bar{x},0;t) &=& \exp{\left[a^2 t
        \frac{1-\bar{x}}{2\bar{x}}\right]}\, q_a(\bar{x})\,,\nn\\ 
\widetilde{H}^a(\bar{x},0;t) &=& \exp{\left[a^2 t
        \frac{1-\bar{x}}{2\bar{x}}\right]}\, \Delta q_a(\bar{x})\,,
\label{gpd}
\ea
where $q(x)$ and $\Delta q(x)$ are the usual unpolarized and polarized
parton distributions in the proton. $a$, the transverse size of the
proton, is the only free parameter and even it is restricted to the
range of about 0.8 to 1.2 $\gev^{-1}$ for a realistic proton. Note
that $a$ mainly refers to the lowest Fock states of the proton
which, as phenomenological experience tells us, are rather compact.
The model (\ref{gpd}) is designed for large $-t$. Hence, forced by the
Gaussian in (\ref{gpd}), large $x$ is implied, too. Despite of this the
normalization of the model GPDs at $t=0$ is correct. 

The model (\ref{gpd}) can be motivated by overlaps of light-cone wave
functions. As has been shown \cite{DFJK1,DFJK3,bro01} GPDs possess a
representation in terms of such overlaps. Assuming a Gaussian
$k_\perp$ dependence for the $N$-particle Fock state wave function
\be
\Psi_N = \Phi_N(x_1,\cdots x_N)\, \exp{\left[-a_N^2 
                           \sum_{i=1}^N k_{\perp i}^2/x_i\right]} \,,
\label{wf}
\ee  
which is in line with the central assumption of the handbag approach
of restricted $k_{\perp i}^2/x_i$, necessary to achieve factorization
of the amplitudes into soft and hard parts, and assuming further $a_N=a$
for all $N$ in order to simplify matters, each overlap provides the
Gaussian appearing in (\ref{gpd}). The remainder of the overlaps
summed over all $N$ is just the Fock state representation of the parton
distribution \cite{bro80}. Thus, there is no need to specify the full 
$x$ dependence of the light-cone wave function in order to arrive at 
(\ref{gpd}). Note that $\Phi_N$ may depend on quark masses.   

The simple model (\ref{gpd}) may be improved in various ways. For
instance, one may treat the lowest Fock states explicitly \cite{DFJK1}, 
take into account the evolution of the GPDs \cite{vogt} or improve the 
parameterization in such a way that it also holds for small $x$
\cite{vander}. One may also consider wave function with a power-law
dependence on ${\bf k}_\perp$ instead of the Gaussian in (\ref{wf}) 
\cite{pauli}.
 
\begin{figure}[t]
\begin{minipage}{0.46\textwidth}
\begin{center} 
\includegraphics[width=3.0cm,bbllx=90pt,bblly=30pt,bburx=590pt,
bbury=635pt,angle=-90,clip=true]{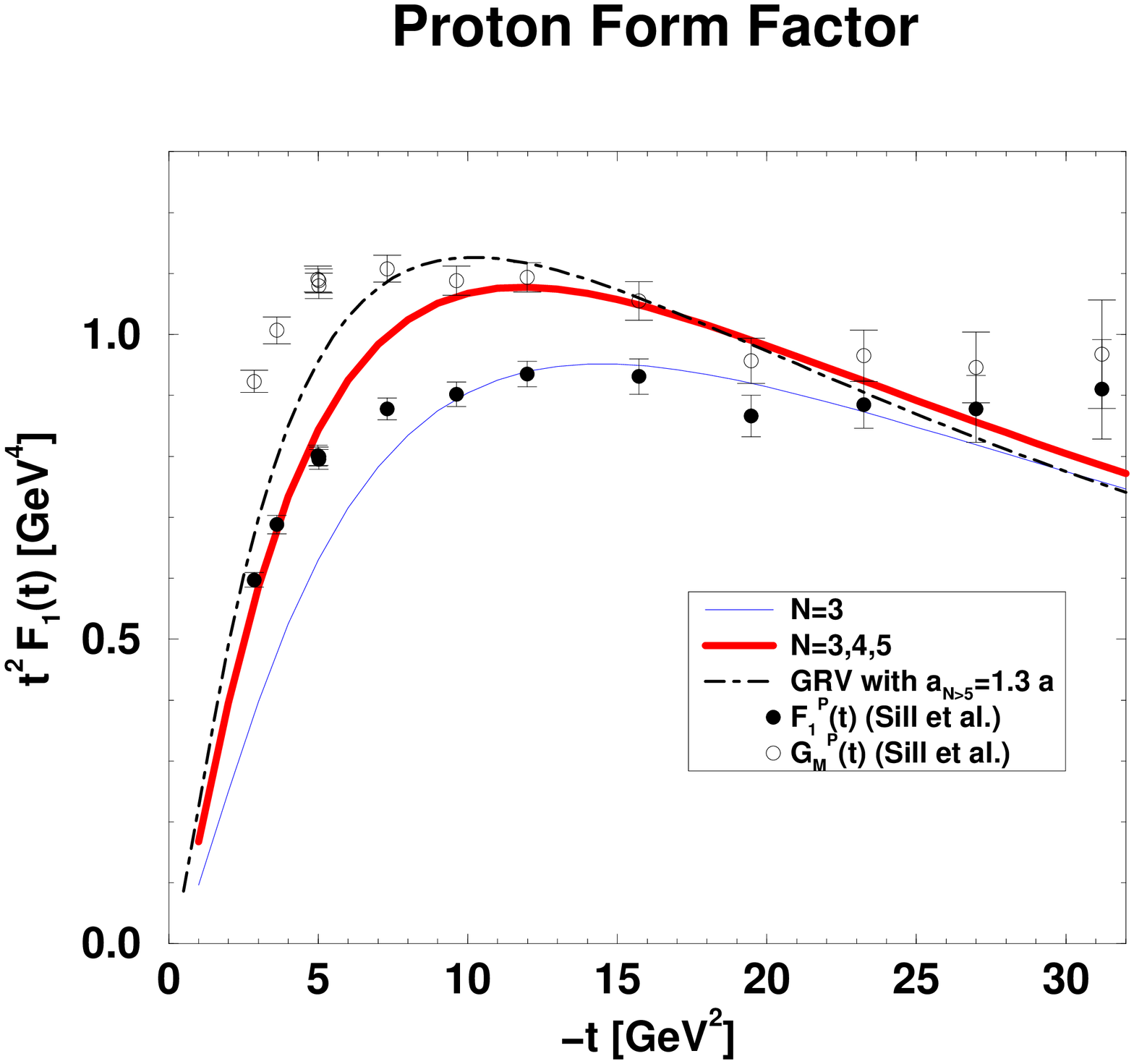}
\end{center}
\end{minipage}
\begin{minipage}{0.46\textwidth}
\begin{center}
\includegraphics[width=4.5cm,bbllx=27pt,bblly=47pt,bburx=398pt, 
bbury=295pt,clip=true]{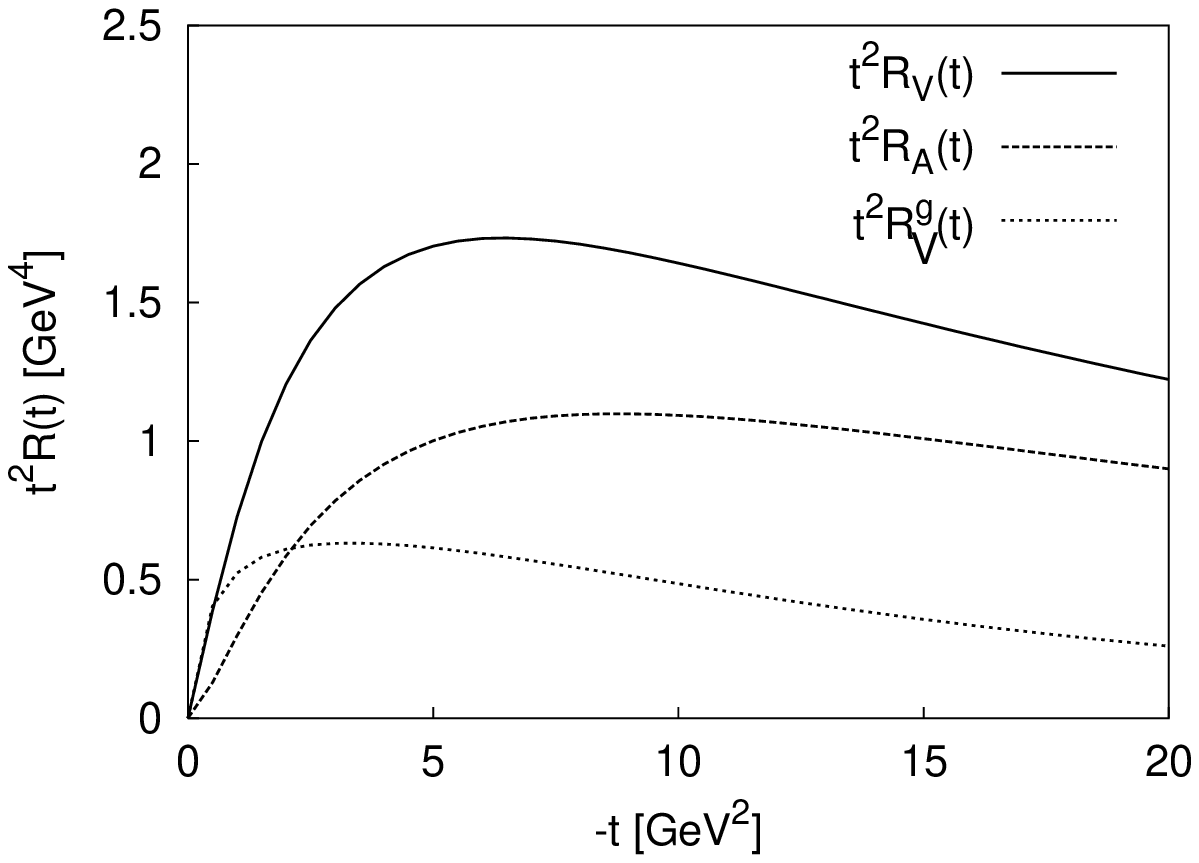}
\end{center}
\end{minipage} 
\caption{Predictions for the Dirac form factor of the proton (left)
and for the Compton form factors (right)\protect\cite{DFJK1}. Data
are taken from Ref.\ \protect\cite{sill}.}
\label{fig:form}
\end{figure}

From the GPDs one can calculate the various form factors by taking
appropriate moments, e.g.\
\be
F_1=\sum_q e_q \int_{-1}^1 d\bar{x} H^q(\bar{x},0;t)\,,\quad
R_V=\sum_q e_q^2 \int_{-1}^1 \frac{d\bar{x}}{\bar{x}} H^q(\bar{x},0;t)\,.
\label{formfactors}
\ee
Results for the form factors are shown in Fig.\ \ref{fig:form}. Obviously,  
as the comparison with experiment \cite{sill} reveals the model GPDs 
work quite well in the case of the Dirac form factor \cite{DFJK1}. The 
scaled form factors $t^2 F_1$ and $t^2 R_i$ exhibit broad maxima which
mimick dimensional counting in a range of $-t$ from, say, $3$ to about
$20\,\gev^2$. For very large values of $-t$, well above 
$100\,\gev^2$, the form factors turn gradually into a $\propto 1/t^4$ 
behaviour; this is the region where the leading-twist contribution
takes the lead. The position of the maximum of a scaled form factor is
approximately located at 
\be
t_0 \simeq -4 a^{-2}\, \left\langle \frac{1-x}{x}\right\rangle^{-1}_{F(R)}\,.
\label{max-pos}
\ee\
The mildly $t$-dependent mean value $\langle (1-x)/x\rangle$ has
a value of about $1/2$.
 
The Pauli form factor $F_2$ and its Compton analogue $R_T$ contribute
to proton helicity flip matrix elements and are related to the GPD $E$
\be
F_2=\sum_q e_q\int_{-1}^1 d\bar{x} E^q(\bar{x},0;t)\,,\quad
R_T=\sum_q e_q^2\int_{-1}^1 \frac{d\bar{x}}{\bar{x}} E^q(\bar{x},0;t)\,.
\ee
The overlap representation of $E$ \cite{DFJK3} involves components of
the proton wave functions where the parton helicities do not sum up to
the helicity of the proton. In other words, parton configurations with
non-zero orbital angular momentum contribute to it. A simple ansatz
for a proton valence Fock state wave function that involves orbital
angular momentum is 
\be
\Psi_3^- \sim \sum \frac{{\bf k}_{\perp i}}{\sqrt{x_i}} \exp{[-a_-^2
\sum {k^2_{\perp i}}/{x_i}]}\,.
\ee
Evaluating the overlap contributions to $F_2$ and $R_T$ from this wave
function and from (\ref{wf}), one finds 
\be
R_T/R_V\,,\;\; F_2/F_1\, \propto  m/\sqrt{-t}
\label{ratio}
\ee
rather than $\propto m^2/t$. (\ref{ratio}) is in agreement with the 
recent JLab measurement \cite{gayou} while the SLAC data \cite{slac} 
are rather compatible with a $\propto m^2/t$ behaviour. The new
experimental results on $F_2/F_1$ have been discussed in the same
spirit as here in Ref.\ \cite{ralston}. Clearly, more phenomenological
work on $E$, $F_2$ and $R_T$ is needed.  

For an estimate of the size of $R_T$ one may simply assume
that $R_T/R_V$ roughly behaves as its electromagnetic counter part
$F_2/F_1$. Hence, 
\be
\kappa_T= \frac{\sqrt{-t}}{2m} \frac{R_T}{R_V} \simeq
                   \frac{\sqrt{-t}}{2m} \frac{F_2}{F_1} 
\ee
has a value of 0.37 \cite{gayou}.
\section{Results for Compton scattering}
\begin{figure}[t]
\begin{center}
\includegraphics[width=4.7cm,bbllx=45pt,bblly=55pt,bburx=320pt, 
bbury=250pt,clip=true]{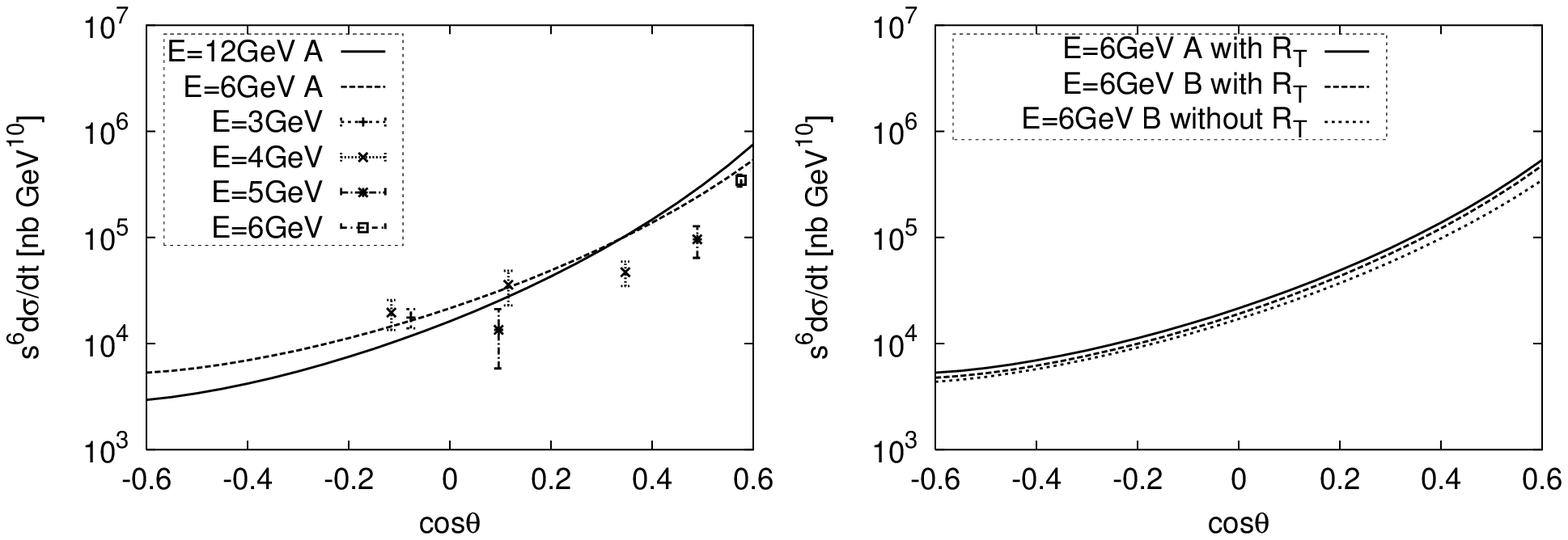}
\includegraphics[width=4.7cm,bbllx=33pt,bblly=47pt,bburx=400pt, 
bbury=295pt,clip=true]{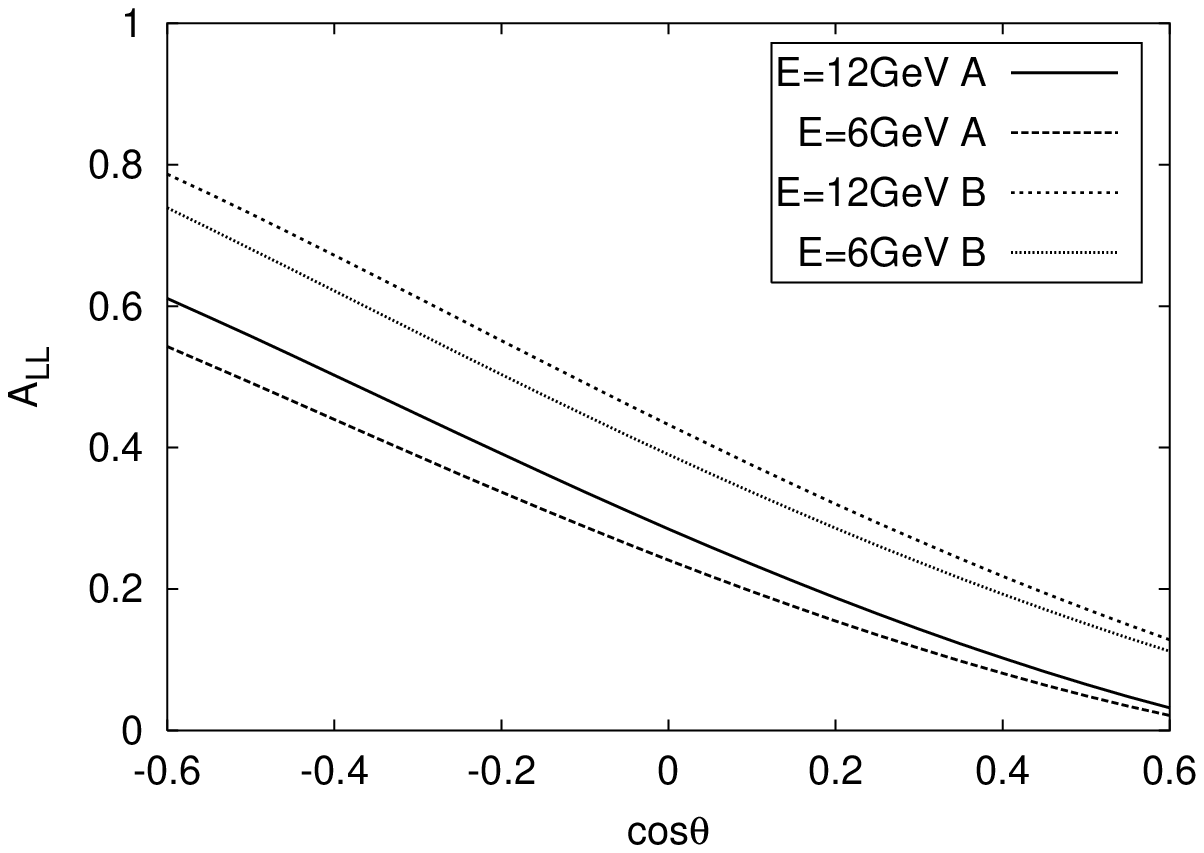} 
\caption{Predictions for the Compton cross section (left) and for the
helicity correlation $A_{LL}$ (right). NLO corrections and the tensor 
form factor are taken into account (scenario A) \protect\cite{hkm}, 
in scenario B they are neglected.  Data are taken from Ref.\
\protect\cite{shupe}.} 
\label{fig:cross}
\end{center}
\end{figure}

I am now ready to discuss results for Compton scattering. The cross
section reads 
\ba
\frac{d\sigma}{dt} &=& \frac{d\hat{\sigma}}{dt} \left\{ \frac12 [
R_V^2(t)(1+\kappa_T^2) + R_A^2(t)] \right.\nn\\
&&\left. \quad  - \frac{us}{s^2+u^2} [R_V^2(t)(1+\kappa_T^2) 
                - R_A^2(t)]\right\} + O(\alpha_s)\,,
\ea
where $d\hat{\sigma}/dt$ is the Klein-Nishina cross section for
Compton scattering of point-like spin-1/2 particles. This cross section
is multiplied by a factor that describes the structure of the proton
in terms of the three form factors. The predictions from the handbag
are in fair agreement with experiment \cite{shupe}, see Fig.\ \ref{fig:cross}.  
The approximative $s^6$-scaling behaviour is related to the broad 
maximum at about $8 \gev^2$ the form factors exhibit, see (\ref{max-pos}). 
Clearly, more accurate date are needed for a detailed comparison. The 
JLab will provide such data soon.

Another interesting observable for Compton scattering is the helicity
correlation, $A_{LL}$,  between the initial state photon and proton
or, equivalently, the helicity transfer, $K_{LL}$, from the incoming
photon to the outgoing proton. From the handbag approach one obtains
\cite{hkm,DFJK2} 
\be
A_{LL}=K_{LL}\simeq \hat{A}_{LL} \frac{R_A}{R_V} +
                                      O(\kappa_T,\alpha_s,\beta)\,,
\ee  
where $\hat{A}_{LL}$ is the corresponding observable for $\gamma q\to
\gamma q$
\be 
\hat{A}_{LL}= \frac{s^2 - u^2}{s^2 + u^2}\,.
\label{sub-all}
\ee
The subprocess observable is diluted by the ratio of the form factors 
$R_A$ and $R_V$ as well as by other corrections but its shape 
essentially remains unchanged. The predictions 
for $A_{LL}$ from the leading-twist approach drastically differ from 
the ones shown in Fig.\ \ref{fig:cross}. For $\theta \lsim 110^\circ$  
negative values for $A_{LL}$ are obtained for all but one 
examples of distribution amplitudes. The diquark model\cite{kroll}, a
variant of the leading-twist approach, also leads to a negative value 
for $A_{LL}$. The JLab E99-114 collaboration \cite{nathan} has
presented a first measurement of $A_{LL}$ at a c.m.s. scattering angle
of $120^\circ$ and a photon energy of $4.3 \gev$. This still
preliminary data point is in agreement with the prediction from the
handbag, the leading-twist calculations fails badly. A measurement of 
the angular dependence of $A_{LL}$ would be highly welcome for
establishing the handbag approach \footnote{Note, however, that, over
a wide range of scattering angles, a Regge model leads to
very similar predictions for $A_{LL}$ as the handbag \cite{laget}.}. 
For predictions of other polarization observables for Compton
scattering I refer to Refs.\ \cite{hkm,DFJK2}.
\section{Other applications of the handbag mechanism}
The handbag approach has been applied to several other high-energy
wide-angle reactions. Thus, as shown in Ref.\ \cite{DFJK2}, the
calculation of real Compton scattering can be straightforwardly  
extended to {\it virtual Compton scattering} provided $Q^2/-t \ll 1$.
{\it Elastic hadron-hadron scattering} can be treated as well \cite{DFJK2}. 
Details have not yet been worked out but it has been shown that form
factors of the type discussed in Sect.\ \ref{sect:model} control 
elastic scattering, too. The experimentally observed 
scaling behaviour of these cross sections can be attributed to the
broad maximum the scaled form factors show, see Fig.\ \ref{fig:form}.

\begin{figure}[t]
\begin{center}
\includegraphics[width=4.3cm]{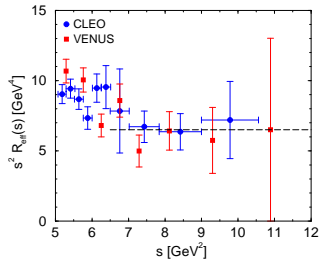} 
\caption{The scaled annihilation form factor $s^2|R_A^p|$
as extracted from the data of Refs.\ \protect\cite{CLEO,VENUS}. The dashed
line represents a fit to the data above \protect$6.5\,\gev\,^2$.}
\label{tl-ff}
\end{center}
\end{figure}

The time-like processes {\it two-photon annihilations into pairs of
mesons or baryons} can also be calculated, the arguments for handbag
factorization hold here as well as has recently been shown in Refs.\
\cite{DKV2,DKV3}, see also the talk by Weiss \cite{weis}.
The cross section for the production of baryon pairs read
\ba
\frac{d\sigma}{dt}\,(\,\gamma\gamma\,\to\,\, B\ov{B}\,) &=& 
      \frac{4\pi\alpha^2_{\rm elm}} 
        {s^2 \sin^2 \theta} \Big\{ \big|R_A^B(s)+ R_P^B(s)\big|^2\nn\\
                  &+& \cos^2\theta\,\big|R_V^B(s)\big|^2\,+\, 
                     \frac{s}{4m^2}\,\big| R_P^B(s)\big|^2 
                                       \Big\}\,.
\ea
The form factors represent integrated $B\ov{B}$ distribution amplitudes 
$\Phi_{B\ov{B}\,i}$ which are time-like versions of GPDs at a time-like
skewness of $1/2$. They read ($i=V,\,A,\,P$)
\be 
R^B_i(s)\,=\, \sum_q\, e_q^2\, F_i^{B q}(s)\,, 
                                      \hspace*{1cm} 
F_i^{B q}(s) = \int_0^1 dz\, \Phi^q_{B\ov{B}\,i}(z,\zeta=1/2,s)\,.
\ee
The form factors have not been modeled by us, they are extracted from
the measured intergrated cross sections. The result for the
effective form factor for $\gamma\gamma\to p\ov{p}$, being a
combination of the dominant axial vector form factor and the
pseudoscalar one, is shown in Fig.\ \ref{tl-ff}. The form factors
behave similar to the magnetic one, $G_M(s)$, in the time-like region and
have the same size as it within about a factor of two \cite{e760}.  
A characterisic feature of the handbag is the $q\ov{q}$ intermediate
state implying the absence of isospin-two components in the final
state. A consequence of this property is 
\be
\frac{d\sigma}{dt}(\gamma\gamma\to \pi^0\pi^0) = 
                 \frac{d\sigma}{dt}(\gamma\gamma\to \pi^+\pi^-)\,,
\ee
which is independent of the soft physics input and is, in so
far, a hard prediction of the handbag approach. 
The absence of the isospin-two components combined with flavor
symmetry allows one to calculate the cross section for other $B\ov{B}$
channels using the form factors for $p\ov{p}$ as the only soft physics
input. It is important to note that the leading-twist mechanism has
difficulties to account for the size of the cross sections
\cite{farrar} while the diquark model \cite{diquark} is in fair
agreement with experiment for $\gamma\gamma\to B\ov{B}$. 

{\it Photo- and electroproduction of mesons} have also been discussed
within the handbag approach \cite{hanwen} using, as in deep virtual
electroproduction \cite{dvem}, a leading-twist mechanism for the
generation of the meson. It turns out, however, that the  
photoproduction cross section is way below experiment. The reason for 
this failure is not yet understood. Either the vector meson dominance
contribution is still large or the leading-twist generation of the meson
underestimates the handbag contribution. Despite of this the handbag
contribution to photo-and electroproduction has several interesting
properties which perhaps survive an improvement of the approach.  For
instance, the helicity correlation $\hat{A}_{LL}$ for the subprocess
$\gamma q \to \pi q$ is the same as 
for $\gamma q \to \gamma q$, see (\ref{sub-all}). $A_{LL}$ for the
full process is diluted by form factors similar to the case of
Compton scattering. Another result is the ratio of the production of
$\pi^+$ and $\pi^-$ which is approximately given by
\be
\frac{d\sigma(\gamma n\to \pi^- p)}{d\sigma(\gamma p\to \pi^+ n)} \simeq
\left[\frac{e_d u + e_u s}{e_u u + e_d s}\right]^2\,.
\ee
\section{Summary}
I have reviewed the theoretical activties on applications of the
handbag mechanism to wide-angle scattering. There are many interesting
predictions still awaiting their experimental examination. At this
workshop many new, mainly preliminary data for wide-angle scattering
from JLab have been presented, more data will come soon. There are
first hints that the handbag mechanism plays an important
role. However, before we can draw firm conclusions we have to wait
till the data have been finalized. For the kinematical situation
available at JLab substantial corrections to the handbag contribution
are to be expected. This may render a detailed quantitative comparison
between theory and experiment difficult. 
\vskip\baselineskip {\it Acknowledgments.}  It is a pleasure to thank
the organizers of this workshop Carl Carlson, Jean-Marc Laget, Anatoly
Radyushkin, Paul Stoler and Bogdan Wojtsekhowski for inviting me to
attend this interesting and well organized workshop at Jefferson Lab.

\end{document}